\documentclass[letterpaper]{article} 
\usepackage[]{aaai2026}  
\usepackage{times}  
\usepackage{helvet}  
\usepackage{courier}  
\usepackage{amsfonts}
\usepackage{amsmath}
\usepackage[hyphens]{url}  
\usepackage[table,xcdraw]{xcolor}
\definecolor{mydarkblue}{RGB}{0,0,180} 
\usepackage{booktabs}
\usepackage{siunitx}
\usepackage{subcaption}
\usepackage{multirow}
\usepackage{graphicx}
\usepackage{placeins}
\usepackage{graphicx} 
\usepackage{natbib}  
\usepackage{caption} 
\usepackage{algorithm}
\usepackage{algorithmic}

\makeatletter
\@ifundefined{isChecklistMainFile}{
  \newif\ifreproStandalone
  \reproStandalonetrue
}{
  \newif\ifreproStandalone
  \reproStandalonefalse
}
\makeatother

\ifreproStandalone
\usepackage{times}
\usepackage{helvet}
\usepackage{courier}
\usepackage{xcolor}
\usepackage{newfloat}
\usepackage{listings}
\DeclareCaptionStyle{ruled}{labelfont=normalfont,labelsep=colon,strut=off} 
\floatstyle{ruled}
\newfloat{listing}{tb}{lst}{}
\floatname{listing}{Listing}
\title{Evolvable Graph Diffusion Optimal Transport with Pattern-Specific Alignment for Brain Connectome Modeling}
	\author{Xiaoqi Sheng$^{1}$, Jiawen Liu$^{2}$, Jiaming Liang$^{2}$, Yiheng Zhang$^{4}$, Hongmin Cai$^{3,*}$.\thanks{$^*$All correspondence should be addressed to Hongmin Cai(Email: hmcai@scut.edu.cn).}}
\affiliations{
$^1$School of Future Technology, South China University of Technology, Guangzhou, Guangdong, China \\ $^2$School of Computer Science and Technology, South China University of Technology, Guangzhou, Guangdong, China \\$^3$School of Future Technology, South China University of Technology, Guangzhou, Guangdong, China \\$^4$School of Computer Science and Technology, Sichuan University, Chengdu, China
}
\usepackage{bibentry}

\begin{document}

\maketitle

\begin{abstract}

Network analysis of human brain connectivity indicates that individual differences in cognitive abilities arise from neurobiological mechanisms inherent in structural and functional brain networks.
Existing studies routinely treat structural connectivity (SC) as optimal or fixed topological scaffolds for functional connectivity (FC), often overlooking higher-order dependencies between brain regions and limiting the modeling of complex cognitive processes. Besides, the distinct spatial organizations of SC and FC complicate direct integration, as naïve alignment may distort intrinsic nonlinear patterns of brain connectivity.
In this study, we propose a novel framework called Evolvable Graph Diffusion Optimal Transport with Pattern-Specific Alignment (EDT-PA), designed to identify disease-specific connectome patterns and classify brain disorders. To accurately model high-order structural dependencies, EDT-PA incorporates a spectrum of evolvable modeling blocks to dynamically capture high-order dependencies across brain regions. Additionally, a Pattern-Specific Alignment mechanism employs optimal transport to align structural and functional representations in a geometry-aware manner.
By incorporating a Kolmogorov–Arnold network for flexible node aggregation, EDT-PA is capable of modeling complex nonlinear interactions among brain regions for downstream classification. Extensive evaluations on the REST-meta-MDD and ADNI datasets demonstrate that EDT-PA outperforms state-of-the-art methods, offering a more effective framework for revealing structure–function misalignments and disorder-specific subnetworks in brain disorders. The project of this work is released via this link.

\end{abstract}


\section{1\quad Introduction}

Accumulating neuroscience evidence identifies structural damage and functional reorganization, which show marked inter-individual variability, as major pathological manifestations of brain disorders~\cite{zhang2011altered}. Furthermore, current pathophysiological models have shifted from emphasizing localized brain pathology to investigating structural and functional interactions across distributed neural networks~\cite{10233896}. Crucially, modern neuroimaging advances allow the construction of graph-theoretical brain connectomes, enabling comprehensive and efficient in vivo characterization of structural connectivity (SC) and functional connectivity (FC) networks~\cite{fornito2015connectomics}.
Specifically, diffusion magnetic resonance imaging (dMRI) noninvasively reconstructs SC by mapping white matter fiber tracts that constitute anatomical pathways for neural information transfer, whereas functional magnetic resonance imaging (fMRI) identifies FC via statistical correlations of neural activity, reflecting dynamic integration processes across distributed brain regions. Existing studies ~\cite{10233896,dan2023uncovering} have demonstrated that SC correlates with FC
 at the group level, underscoring their complementary value in enhancing classification accuracy. However, a primary challenge is that SC reflects only anatomical connections, whereas FC represents interactions that may occur independently of direct anatomical links, resulting in an incomplete correspondence between the two~\cite{popp2024structural}. Additionally, despite advances in data-driven methods for brain network analysis~\cite{11059325}, effectively encoding higher-order topological features to identify clinically relevant biomarkers for neurological disorders remains a significant challenge. Therefore, accurately modeling the topological properties of SC and FC is critical for understanding complex cognitive functions and brain behaviors.

Recently, driven by the need for automated diagnosis~\cite{bahrami2023brain}, various deep learning approaches based on brain connectivity data have been developed to enable precise analysis of disease-specific topological alterations and cognitive impairment~\cite{wang2025brainmap,ma2023brainnet,pang2023slim}.
Among these, Graph Convolutional Networks (GCNs)~\cite{parisot2018disease,9933896,8944082} have emerged as particularly powerful for brain network analysis, owing to their inherent capacity to model the topological relationships between SC-FC. However, existing GCNs primarily focus on distinguishing brain region features, often overlooking critical topological information for modeling brain propagation patterns.
One strategy is known as the guided method, where one connectivity modality (e.g., SC) directly guides the estimation of another (e.g., FC). For example, Pinotsis et al. ~\cite{pinotsis2013anatomical} pioneered a guided connectivity approach, leveraging SC to establish theoretical relationships between graph topological properties and simulated functional dynamics. By leveraging multiple generative adversarial networks (GANs), Tan et al. ~\cite{tan2025connectome} introduced a framework for cross-modal connectivity synthesis and translation, achieving substantial improvements in brain disorder classification accuracy. Bian et al.'s ~\cite{10233896} topology-aware GCN framework integrates homology features from SC to constrain FC estimation, enhancing sensitivity to pathological microstructural changes. However, a major limitation of these methods is their focus solely on direct connectivity, neglecting indirect connectivity at broader scales, which significantly diminishes the reliability of diagnostic outcomes. This limitation arises from the fact that information transmission between brain regions is characterized by both strong local low-order connections and efficient high-order connections~\cite{stam2014modern}. Several studies have applied Transformer architectures to the graph domain to address the aforementioned challenges~\cite{10518112,10645714}. These methods typically employ specialized positional embedding strategies that integrate FC and SC information to optimize the computation of global features across brain regions. However, given that the correlations between functional and structural pathways are not linear, traditional graph transformers struggle to accurately reflect the underlying biological mechanisms~\cite{10182318}. Therefore, joint combinatorial reasoning of functional and structural connectomes should be incorporated into graph-based modeling.


\begin{figure}[!t]
    \centering      
    \includegraphics[width=0.475\textwidth]{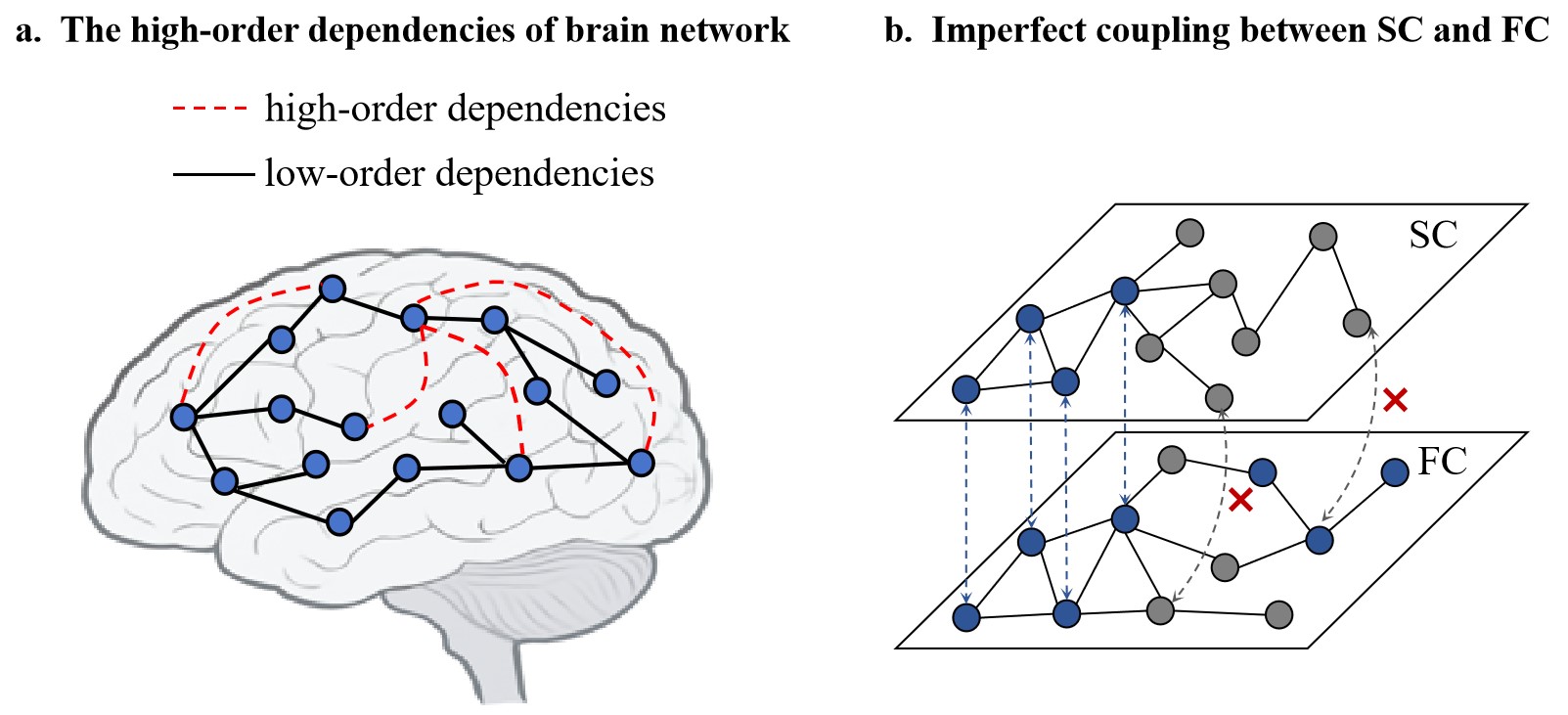}
    \caption{An illustration of the challenges in integrating SC and FC, highlighting two key issues: 
(a) High-order dependencies: Red dashed lines indicate high-order dependencies arising from indirect pathways or coordinated activity across multiple brain regions.
(b) Imperfect coupling between SC and FC: Differences in spatial distribution and organization between SC and FC.
 }
    \label{fig: challeng}
    \vspace{-0.255cm}
\end{figure}


The above analysis demonstrates that successful SC and FC analysis methods effectively capture both the intrinsic connectivity patterns and the disease-specific discriminatory information.
However, recent studies still rely on the assumption that \textbf{the structural brain network can serve as the optimal graph topology for the functional network}~\cite{tan2025connectome}.  This premise introduces two inherent challenges that could significantly compromise downstream disease pattern classification performance:
(1) \textbf{\textit{Neglect of high-order dependencies}}. As shown in Fig.~\ref{fig: challeng}a, restricting information propagation strictly to SC-defined edges fails to account for indirect functional interactions and distributed neural coordination, limiting the model’s capacity to capture higher-order cognitive processes.
(2)\textbf{ \textit{Imperfect coupling between SC and FC}}. As shown in Fig.~\ref{fig: challeng}b, SC and FC exhibit significant statistical discrepancies in spatial distribution and organization. Directly integrating them may distort the nonlinear spatial structure of brain networks, ultimately compromising the generalizability and interpretability of the resulting models.

In light of the above-mentioned issues, in this study, we introduce an Evolvable Graph Diffusion–Optimal Transport with Pattern-Specific Alignment (\textbf{EDT-PA}) to classify brain disorders. The framework is built upon two key components:
(1) \textbf{EBCM} (Evolvable Brain Connectome Modeling), which employs an innovative iterative graph diffusion optimization strategy to disentangle complex pathways within SC, generating an adaptive SC adjacency matrix with higher-order dependency encoding; and
(2) \textbf{PSSA} (Pattern-Specific Structure–Function Alignment), which leverages an optimal transport strategy to develop an edge-aware graph encoder that bridges high-order SC and FC characteristics, enhancing model interpretability while maintaining strong generalization capacity.
Extensive experiments on two benchmark brain network datasets, REST-meta-MDD and ADNI, validate the superiority of EDT-PA. Compared with state-of-the-art (SOTA) methods, our framework achieves improvements of 5.4\% and 12.3\% in crucial accuracy metrics, while maintaining robust interpretability and generalization across tasks. 
In summary, our contributions are as follows:
\begin{itemize}
        \item \underline{\textbf{\textit{Innovation.}}}
        An evolvable graph diffusion optimal transport method is proposed for brain connectome modeling, dynamically capturing interregional dependencies to integrate structural and functional networks and enhance disease-specific pattern identification.

        \item \underline{\textbf{\textit{Architecture.}}}
        A comprehensive end-to-end joint analysis framework, EDT-PA, has been designed. This method integrates evolvable modeling modules to capture high-order inter-regional dependencies, enabling precise characterization of disease-related network abnormalities.

        \item \underline{\textbf{\textit{Validation.}}}
    Extensive experiments on benchmark datasets demonstrate the method's superiority over state-of-the-art approaches in disease classification, robustly identifying discriminative disease regions.
    \end{itemize}

\section{2\quad Preliminary}
The brain connectome graph, constructed from fMRI, sMRI, and DTI data, can be formally represented as a graph $G = (\mathcal{V}, \mathcal{E},A,X)$. Here, the $\mathcal{V} =\{v_i \mid i=1,\ldots,N\}$ denotes the set of nodes corresponding to brain regions of interest (ROIs),  while the $\mathcal{E} = \{a_{i,j} \mid (v_i, v_j) \in \mathcal{V} \times \mathcal{V}\}$ represent connections between brain regions' nodes. The adjacency matrix $A \in \mathbb{R}^{N \times N}$ encodes the strength of inter-nodal connections, and the node feature matrix $X\in \mathbb{R}^{N \times d}$ is derived from Pearson correlation coefficients~(PCC) computed from blood-oxygen-level-dependent (BOLD) signals in fMRI data. Each brain graph is associated with a categorical label $y$, indicating the subject’s clinical or physiological condition.

\section{3\quad Methodology}

\begin{figure*}[!t]
    \centering      \includegraphics[width=1.0\textwidth]{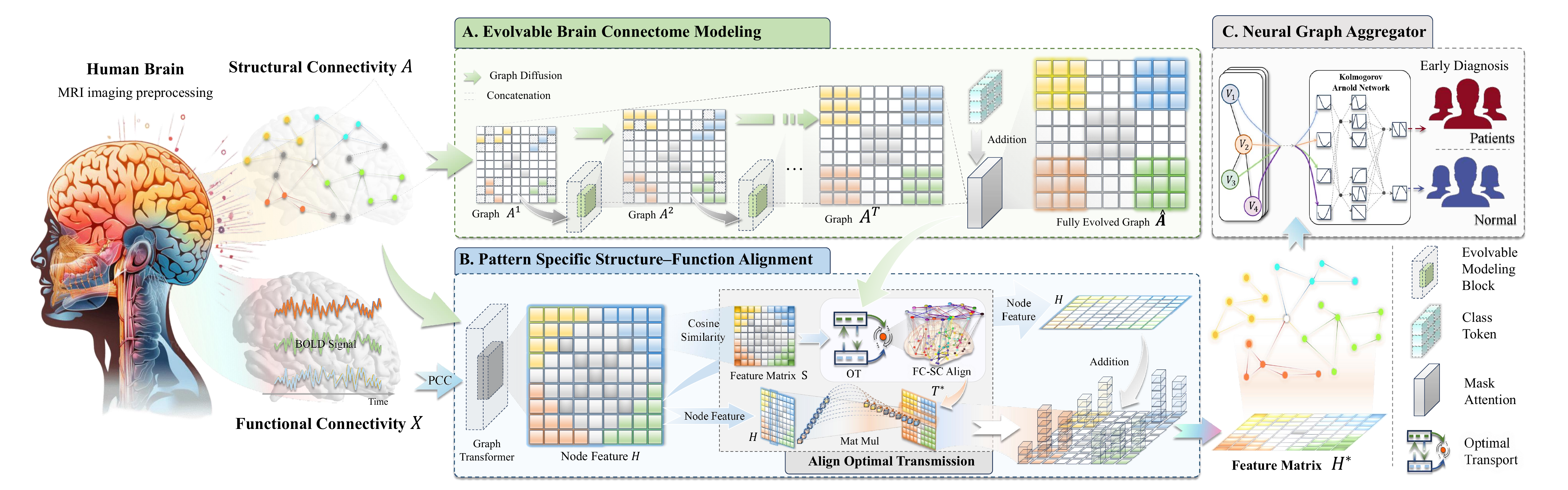}
    \caption{Architecture of the proposed EDT-PA for brain connectome modeling }
    \label{fig: overview}
    \vspace{-0.205cm}
\end{figure*}

The proposed EDT-PA framework, as illustrated in Fig.~\ref{fig: overview}, integrates three core modules to achieve effective brain network analysis: (1) an \textbf{Evolvable Brain Connectome Modeling module} that progressively refines structural connectivity through multi-step graph diffusion and class-aware Transformers, (2) a \textbf{Pattern Specific Structure-Function Alignment module} that establishes precise neurobiological correspondence via attention-guided fusion and optimal transport-based matching, and (3) a \textbf{Neural Graph Aggregator} that models intricate regional interactions through Kolmogorov-Arnold Networks for robust downstream classification. By effectively bridging FC and SC representations, the model enables more comprehensive neural interaction analysis for improved brain disorder classification.


\subsection{3.1\quad Evolvable Brain Connectome Modeling}
Accurate representation of brain structural connectivity is crucial for downstream clinical tasks~\cite{9942309}. However, raw structural connectomes are typically represented as sparse, noisy symmetric matrices, which hinder existing algorithms from capturing higher-order interactions and functional dependencies across distant brain regions~\cite{10182318}.
To address these limitations, EDT-PA develops an EBCM pipeline. The proposed methodology advances graph diffusion processes via multiple hierarchical receptive fields, while a class-conditional Transformer architecture provides adaptive learning of spatiotemporal correlations. The diffusion steps of EBCM are formulated as follows:
\vspace{-0.105cm}
\begin{equation}
\resizebox{.9\linewidth}{!}{$
A^{(t+1)} = \mathcal{T}\big(\alpha SA^{(t)}S^{T} + (1-\alpha)A\big),t= 0,1,\ldots,T-1$} 
\end{equation}
where, $S~=~D^{-\frac{1}{2}} A D^{-\frac{1}{2}}$ represents the diffusion operator, and D is a diagonal matrix
with elements~$\quad D_{ii}=\sum_{j=1}^{N} A_{ij}$~\cite{bai2017regularized}.
The hyperparameter $\alpha$ controls the trade-off between the diffused structure and the original graph topology. To capture the high-order structural dependencies inherent in brain networks, a Transformer model $\mathcal{T}$ is integrated at each diffusion step. The self-attention mechanism of the Transformer explicitly models high-order connectomic interactions, addressing the limitations imposed by local neighborhood constraints in traditional graph diffusion.  This integration significantly enhances the model's capacity to represent complex, large-scale brain organization.
The process generates a sequence $\mathcal{A} = \{A^{(1)}, A^{(2)}, \dots, A^{(T)}\}$, encoding progressive structural relationships across brain regions.

Another central problem in neuroscience is extracting unique connectivity patterns from the structural connectome that are associated with brain diseases.
To this end, a class token $e_y$ is computed and incorporated into the modeling of the fully evolved graph.
Specifically, $e_y$ is obtained via an index-based lookup operation \( e_y = \mathcal{M}(E[Y]) \), where \( \mathcal{M}: \mathbb{R}^{1 \times N^2} \rightarrow \mathbb{R}^{N \times N} \) representing the reshape operation, and \( E \in \mathbb{R}^{C \times N^2} \) is a learnable embedding matrix.

In the absence of accessible class labels during inference, a soft class-query module is introduced to compute a probabilistic class embedding directly from the input features, enabling implicit task-aware conditioning. Formally, given the adjacency matrix $A \in \mathbb{R}^{N\times N}$ of the brain connectome graph $G$, the query-to-class attention is computed as:
\begin{equation}
    \begin{aligned}
    \beta = softmax(\mathcal{M}^{-1}(A)*E^{T}) \\
    e_y = \mathcal{M}(\beta\cdot E)\in \mathbb{R}^{N\times N} 
    \end{aligned}
\end{equation}
in which $\mathcal{M}^{-1}: \mathbb{R}^{N\times N} \rightarrow \mathbb{R}^{1 \times N^{2}}$ is the reverse operation of $\mathcal{M}$.
The soft class token is then appended to the structural diffusion sequence to enable task-aware conditioning without requiring explicit class labels during inference. 
Once $e_y$ is obtained, it is added as a global prompt token to the sequence $\mathcal{A}$:
\begin{equation}
    \mathcal{A}^{*} = \{e_y,A^{(1)}, A^{(2)}, \dots, A^{(T)}\}
\end{equation}

In essence, the $\mathcal{A}^{*}$ contains the progressively diffused graph structures along with the class token, which encapsulates disease-specific connectivity patterns. To derive the fully evolved graph from these representations, an autoregressive model is employed to iteratively refine and expand upon $\mathcal{A}^{*}$:


\vspace{-0.205cm}
\begin{equation}
            p\left(\hat{A} \mid \mathcal{A}^{*}\right) \\
     =\prod_{k=1}^K p\left(\hat{A}\mid e_{y}, A^{(1)}, \ldots, A^{(T)}\right)
\end{equation}

To capture the conditional dependency $p\left(\hat{A} \mid \mathcal{A}^{*}\right)$, a Transformer module with mask attention is employed to approximate the underlying distribution. This process produces the final output $\hat{A}$, which serves as the updated structural connectome, enriched with multi-scale awareness and task-specific modulation.


Collectively, the procedure implements an anatomical prior encoding mechanism that simulates neural signal propagation, emphasizing informative pathways through class-aware guidance.

\subsection{3.2~Pattern Specific Structure-Function Alignment}

Through iterative diffusion optimization, the structural connectivity graph gradually evolves, capturing richer high-order dependencies. However, these features are still confined to the brain's SC modality. Therefore, we further employ the PSSA module, which aims to refine the alignment between SC and FC via an optimal transport mechanism, enabling more accurate modality fusion and enhancing the expressiveness of the brain connectome.
Specifically, the structural connectivity matrix $A$ and the node features $X\in\mathbb{R}^{N\times d}$ are first integrated using a Graph Transformer with edge features.

In graph Transformer layer, the feature of a given brain region (node) $x_{i}$
  is concatenated with those of its structurally adjacent regions:
\begin{equation}
    h_{i} = \|(x_i,\{x_{j}\mid j \in \mathcal{N}_{i}\})
\end{equation}
where $\|$ denotes the concatenation and $\mathcal{N}_{i}$ represents the set of neighbors of node $i$. This concatenated representation is then processed by the Transformer module, followed by integration with edge-level connectivity features $\{a_{j}\mid j\in \mathcal{N}_{i}\}$:
  \begin{align}
   &h_{i} = \mathcal{T}_{1}(h_i) ,a_{j} = \mathcal{T}_{2}(a_{j}) \\
   &h_{i} = \sum_{j\in\mathcal{N}_{i}}a_{ij}h_{j}
  \end{align}
where $\mathcal{T}_{1}$ and $\mathcal{T}_{2}$ refer to two distinct Transformer layers, each with independent parameters.
Once the graph embedding $H = \{h_i\mid i=1,\ldots,N\}$ is obtained, cosine similarity is used to compute the similarity between each pair of nodes: 
\begin{equation}
    S = \frac{H\cdot H^{T}}{\|H\| \times \|H\|} \in \mathbb{R}^{N \times N},H \in \mathbb{R}^{N\times nd}
\end{equation}
where $n$ denotes the number of neighboring nodes associated with each node.


Recognizing the complexity and inherent misalignment between SC and FC, EDT-PA introduces a novel optimal transport (OT) strategy that transcends traditional alignment approaches.
Unlike prior works that rely on standard OT to align fixed topologies or embed SC into FC via heuristic rules, our method constructs a transport-aware correspondence that is dynamically informed by both functional similarity $S$ and diffusion-derived structural priors $\hat{A}$. 
For convenience in the subsequent formulas, $\hat{A}$ and $S$ are rewritten as: $\hat{A}=\{a_{i}\}_{i=1}^{N}$ and $S = \{s_{i}\}_{i=1}^{N}$. Next, the discrete empirical distributions $u$ and $v$, defined on the probability spaces $\hat{A}, S \in \Omega$, are presented as follows: 

\begin{equation}
    \mathbf{u}=\sum_{i=1}^N \mu_i \delta_{\boldsymbol{a}_i}, \quad \mathbf{v}=\sum_{j=1}^N \nu_j \delta_{\boldsymbol{s}_j},
\end{equation}
where, $\delta_{\boldsymbol{a}_i}$ and $\delta_{\boldsymbol{s}_j}$ denote Dirac functions centered on $\hat{A}$ and $S$, respectively. The weight vectors $\mu = \{{\mu_{i}\}}_{i=1}^{N}$ and $\mathbf{v}=\{{\mathbf{v}_{j}\}}_{j=1}^{N}$ belong to the N -dimensional simplex, i.e., $\sum_{i=1}^{N}\mu_{i}=1$,
$\sum_{j=1}^{N}\mathbf{v}_{i}=1$.

The alignment is formalized as an entropy-regularized optimal transport problem, which is solved using the Sinkhorn-Knopp algorithm~\cite{10.5555/2999792.2999868}. Specifically, a transport plan $\mathbf{T}^*$ is calculated to minimize the expected transport cost between $\hat{A}\in\mathbb{R}^{N\times N}$ and $S\in\mathbb{R}^{N \times N}$, subject to marginal constraints:
\begin{equation}
       \mathbf{T}^*=\arg \min _{\mathbf{T} \in \mathbb{R}^{N \times N}}\langle \mathbf{T}, C\rangle-\epsilon Z(\mathbf{T}), \text { s.t. } \mathbf{T} \mathbf{1}=\mu, \mathbf{T}^{\top} \mathbf{1}=\nu
\end{equation}
where $C$ is a cost matrix defined by pairwise dissimilarities between 
$A$ and $S$, computed via the cosine distance. $Z(T)=-\sum_{ij}T_{ij}logT_{ij} $ denotes the entropy of the transport plan $T$, and $\epsilon >0$ is a smoothing parameter controlling the regularity of the transport. For the problem Eq.~(11), we have an optimization solution when
$t \rightarrow \infty$:
\begin{equation}
\mathbf{T}^*=\operatorname{diag}\left(\mathbf{a}^t\right) \exp (-\mathbf{C} / \epsilon) \operatorname{diag}\left(\mathbf{b}^t\right)
\end{equation}
in which $t$ is the iteration steps, $a^{t} = \frac{\mu}{exp(-{C}/{\epsilon})}b^{t-1}$ and $b^{t} = \frac{\nu}{exp({C}/{\epsilon})}a^{t}$, with the initialization on $b^{0}=1$.
The stability of the iterative computations is ensured by employing the logarithmic scaling variant of Sinkhorn optimization~\cite{schmitzer2019stabilized}. 
The biologically meaningful transport plan $\mathbf{T}^*$ aligns the intrinsic organization of SC and FC, refining the node features as follows:
\begin{equation}
    H^* = T^{*}H+H
\end{equation}

By embedding this transport mechanism into a modular Graph Transformer framework with explicit edge awareness, we achieve fine-grained, pattern-specific alignment between FC and SC.

\subsection{3.3\quad Neural Graph Aggregator}

Given the complexity and spatial variability of inter-regional brain interactions, we extend the Kolmogorov-Arnold Network~(KAN) ~\cite{liu2024kan} into graph neural networks as a node-level aggregation mechanism in functional brain graphs.


For each node $i$, its refined representation $h^{*}_i$ is updated by jointly considering its current state and the representations of its neighbors $\{h^{*}_j\mid j \in \mathcal{N}(i)\}$ through the KAN aggregation function:
\begin{equation}
    h^{*}_i =KAN(h^{*}_i,\{h^{*}_j\}_{j\in\mathcal{N}_{i}}) = \Phi_{L-1} \circ \cdots \circ \Phi_0(h^{*}_i,\{h^{*}_j\})
\end{equation}
where each transformation $\Phi_{l}$ represents a deep, nonlinear transformation that learns progressively higher-order interactions between the node $i$ and its neighbors $\mathcal{N}_{i}$

Once the node-level feature representations are updated, we proceed to compute a graph-level embedding
$h^{*}_{G}$ by performing a mean readout operation across all node representations in the graph:
\begin{equation}
     h^{*}_G = \frac{1}{|V|} \sum_{i \in V} h_i^{*}
\end{equation}
This graph-level embedding, which captures the global structure of the brain network, is then passed through a multi-layer perceptron (MLP) for final classification: $\hat{y} = \mathcal{F}(h^{*}_G)$. Therein, $\mathcal{F}(\cdot)$ is implemented as an MLP with a softmax output. Given the ground truth label $y$, the loss function of our proposed model is formulated as follows:
\begin{equation}
	\begin{aligned}
		loss = \mathcal{L}_{ce}(y,\hat{y})=-\mathbb{E}_{y}[\operatorname{log}(\hat{y})]
	\end{aligned}
\end{equation}
where $\mathbb{E}$ is the expectation and $\mathcal{L}_{ce}$ represents the cross-entropy loss function.

\section{4\quad Experiment}

\subsection{4.1\quad Data Description}
In this study, a comprehensive evaluation of EDT-PA's effectiveness in brain disease diagnosis is conducted on two publicly available datasets: REST-meta-MDD and ADNI.  Descriptions of these datasets are provided below.

\textbf{REST-meta-MDD}:
The REST-meta-MDD consortium provides standardized sMRI/fMRI data for major depressive disorder (MDD) research, including 781 matched subjects (395 MDD patients and 386 controls). MRI data underwent preprocessing, including motion correction, T1-MRI alignment, SPM12-based MNI normalization, and 5mm FWHM Gaussian smoothing~\cite{dadi2019benchmarking}. Structural connectivity (SC) was derived from Jensen-Shannon Divergence (JSD) of inter-regional gray matter volume similarity~\cite{wang2016single, li2021surface, sebenius2023robust}. The brain was parcellated into 264 regions using the Power264 atlas~\cite{power2011functional}, and functional connectivity (FC) was computed via Pearson correlations of BOLD time series from these ROIs.

\textbf{ADNI}:
The Alzheimer's Disease Neuroimaging Initiative (ADNI) provides a multimodal dataset for Alzheimer's disease (AD) research, including sMRI, fMRI, PET, diffusion imaging, CSF, blood biomarkers, genetic profiles, and cognitive assessments from 203 AD patients and 103 cognitively normal controls (CN), matched for age and sex. Imaging data underwent skull-stripping, with T1-weighted and rs-fMRI co-registered to DTI space using FLIRT~\cite{jenkinson2012fsl}. Rs-fMRI preprocessing included spatial smoothing, slice-timing correction, temporal prewhitening, drift removal, and bandpass filtering (0.01–0.1 Hz). Diffusion data were corrected for eddy currents and processed with MedINRIA~\cite{toussaint2007medinria} for fiber tractography. T1 images were parcellated into 148 cortical regions using the Destrieux Atlas~\cite{destrieux2010automatic} in FreeSurfer~\cite{fischl2012freesurfer}, defining SC nodes. SC matrices were constructed by counting streamlines between regions, and FC was derived from Pearson correlations of BOLD time series.

\begin{table*}[htbp]
    \centering
    \caption{Performance comparison on REST-meta-MDD and ADNI datasets}
    \label{tab:comparison}
    \small
    \renewcommand{\arraystretch}{1.1}
    \setlength{\tabcolsep}{5pt}
    \begin{tabular}{llccccc}
        \toprule
        \textbf{Dataset} & \textbf{Method} & \textbf{ACC}                & \textbf{PRE}                & \textbf{REC}                & \textbf{F1}                 & \textbf{AUC}                \\
        \midrule
        \multirow{11}{*}{REST-meta-MDD}
        & RF              & 0.567$\pm$0.021             & 0.572$\pm$0.024             & 0.567$\pm$0.021             & 0.567$\pm$0.020             & 0.585$\pm$0.023             \\
        & SVM             & 0.552$\pm$0.043             & 0.553$\pm$0.045             & 0.553$\pm$0.043             & 0.550$\pm$0.042             & 0.523$\pm$0.081             \\
        & GCN             & 0.558$\pm$0.030             & 0.463$\pm$0.147             & 0.558$\pm$0.030             & 0.555$\pm$0.033             & 0.567$\pm$0.013             \\
        & GIN             & 0.564$\pm$0.030             & 0.569$\pm$0.025             & 0.564$\pm$0.031             & 0.559$\pm$0.040             & 0.560$\pm$0.038             \\
        & GraphGPS        & 0.577$\pm$0.034             & 0.568$\pm$0.038             & \textbf{0.780$\pm$0.126}    & 0.650$\pm$0.027             & 0.597$\pm$0.054             \\
        & BrainGNN        & 0.544$\pm$0.026             & 0.527$\pm$0.031             & 0.741$\pm$0.226             & 0.598$\pm$0.074             & 0.531$\pm$0.071             \\
        & BrainGB         & \underline{0.727$\pm$0.023} & \underline{0.762$\pm$0.035} & 0.727$\pm$0.020             & \underline{0.718$\pm$0.020} & \underline{0.838$\pm$0.032} \\
        & BrainIB         & 0.636$\pm$0.012             & 0.639$\pm$0.015             & 0.655$\pm$0.028             & 0.643$\pm$0.015             & 0.663$\pm$0.014             \\
        & ATPGCN          & 0.654$\pm$0.013             & 0.660$\pm$0.043             & 0.692$\pm$0.128             & 0.668$\pm$0.046             & 0.690$\pm$0.031             \\
        & BrainGRL        & 0.682$\pm$0.022             & 0.673$\pm$0.050             & 0.738$\pm$0.107             & 0.696$\pm$0.030             & 0.683$\pm$0.038             \\
        & EDT-PA (ours)   & \textbf{0.781$\pm$0.027}    & \textbf{0.787$\pm$0.027}    & \underline{0.771$\pm$0.038} & \textbf{0.778$\pm$0.031}    & \textbf{0.841$\pm$0.045}    \\
        \midrule
        \multirow{11}{*}{ADNI}
         & RF             & 0.613$\pm$0.027             & 0.518$\pm$0.144             & 0.613$\pm$0.027             & 0.499$\pm$0.033             & 0.539$\pm$0.046             \\
         & SVM            & 0.619$\pm$0.060             & 0.614$\pm$0.050             & 0.619$\pm$0.060             & 0.611$\pm$0.051             & 0.590$\pm$0.014             \\
         & GCN            & 0.629$\pm$0.018             & 0.596$\pm$0.026             & 0.629$\pm$0.177             & 0.589$\pm$0.125             & 0.635$\pm$0.015             \\
         & GIN            & 0.642$\pm$0.051             & 0.556$\pm$0.092             & 0.642$\pm$0.051             & 0.575$\pm$0.069             & 0.547$\pm$0.080             \\
         & GraphGPS       & 0.639$\pm$0.037             & 0.666$\pm$0.014             & \underline{0.933$\pm$0.057} & 0.777$\pm$0.028             & 0.544$\pm$0.085             \\
         & BrainGNN       & 0.642$\pm$0.026             & 0.644$\pm$0.028             & \textbf{0.975$\pm$0.015}    & 0.775$\pm$0.017             & 0.557$\pm$0.036             \\
         & BrainGB        & 0.654$\pm$0.061             & 0.623$\pm$0.053             & 0.620$\pm$0.054             & 0.617$\pm$0.055             & 0.635$\pm$0.078             \\
         & BrainIB        & 0.663$\pm$0.016             & 0.722$\pm$0.015             & 0.808$\pm$0.030             & 0.759$\pm$0.014             & 0.623$\pm$0.020             \\
         & ATPGCN         & 0.677$\pm$0.025             & 0.698$\pm$0.014             & 0.909$\pm$0.028             & 0.790$\pm$0.017             & 0.670$\pm$0.025             \\
         & BrainGRL       & \underline{0.719$\pm$0.061} & \underline{0.747$\pm$0.061} & 0.886$\pm$0.046             & \underline{0.809$\pm$0.040} & \underline{0.741$\pm$0.071} \\
         & EDT-PA (ours)   & \textbf{0.842$\pm$0.012}    & \textbf{0.843$\pm$0.013}    & 0.842$\pm$0.012             & \textbf{0.835$\pm$0.015}    & \textbf{0.828$\pm$0.025}    \\
        \bottomrule
    \end{tabular}
\end{table*}

\subsection{4.2\quad Comparison Methods and Settings}

To validate the effectiveness of our proposed EDT-PA model, we compare its performance with a range of classical machine learning classifiers and SOTA graph learning methods on REST-meta-MDD and ADNI. These comparative methods can be broadly categorized into four groups:
\begin{itemize}
    \item \textbf{Baseline models:} Random Forest (RF)~\cite{rigatti2017random} and Support Vector Machine (SVM)~\cite{jakkula2006tutorial}.

    \item \textbf{General Graph Neural Network Methods:} GCNs~\cite{parisot2018disease}, Graph Isomorphism Network (GIN)~\cite{duan2023predicting} and GraphGPS~\cite{rampavsek2022recipe}.

    \item \textbf{Brain Network-Specific Graph Models:} BrainGNN~\cite{li2021braingnn}, BrainGB~\cite{9933896} and BrainIB~\cite{10680255}.
    \item \textbf{Joint SC-FC Modeling Methods:} BrainGRL~\cite{li2022learning} and ATPGCN~\cite{10233896}.
\end{itemize}

In this experiment, the datasets are randomly partitioned into 70$\%$, 10$\%$, and 20$\%$ for training, validation, and testing, respectively. Additionally,  weight $\alpha$ and diffusion steps $T$ are empirically set to $0.3$ and $4$, respectively. Five evaluation metrics are used to assess the algorithm's performance, including accuracy (ACC), recall (REC), precision (PRE), area under the ROC curve (AUC), and F1-score (F1).
To ensure experimental fairness, both EDT-PA and the comparative methods were trained and evaluated under the same setup as previously outlined. The experimental results of all methods are summarized in Table~\ref{tab:comparison}, with the best values for each evaluation metric highlighted in bold and sub-SOTA values underlined.
\subsection{4.3\quad Classification Performance}
The classification results reported in Table~\ref{tab:comparison} show that, across two benchmark datasets, the proposed EDT-PA model demonstrates clear advantages in both accuracy and robustness.
These benefits are particularly significant given the substantial differences between the datasets in terms of sample size, data heterogeneity, and brain network construction methods.

On the REST-meta-MDD dataset, EDT-PA outperforms strong baselines like BrainGNN by dynamically evolving the structural graph and more effectively aligning the functional topology. This results in improvements of 5.4\% and 6.0\% in ACC and F1, respectively, surpassing the performance of the sub-SOTA method. 
Notably, although competing approaches such as GraphGPS and BrainGNN achieve relatively higher recall, their performance is compromised by substantially lower precision and AUC scores. This limitation arises from their dependence on either global attention mechanisms or static node representations, which constrains their generalization capacity and leads to systematic over-prediction of positive cases.

To further evaluate the generalizability of EDT-PA, additional experiments are conducted on the more challenging ADNI dataset. In this severely imbalanced classification task (203 ADs vs. 103 CNs), EDT-PA achieved an accuracy of 84.2\% and an AUC of 82.8\%, significantly outperforming all baseline methods. Models such as ATPGCN and BrainGRL, which integrate SC and FC, demonstrate superior performance over most baselines on the ADNI dataset. However, their effectiveness is limited by the intrinsic constraints of their fusion strategies. Specifically, these models lack mechanisms explicitly designed to address modality heterogeneity and resolve alignment mismatches. Consequently, despite achieving high REC, they exhibit significant overfitting to the positive class, as evidenced by their comparatively lower ACC. In contrast, EDT-PA employs an OT-based alignment strategy that selectively aligns connectivity patterns between structural and functional modalities, rather than enforcing full distribution-level alignment. This targeted strategy mitigates the risk of overfitting the dominant class and enhances the robustness to modality-specific noise. As a result, EDT-PA achieves superior performance in both ACC and REC, demonstrating strong robustness and generalization across heterogeneous neuroimaging data.


\subsection{4.4\quad Ablation Study}

\begin{figure}[!t]
    \includegraphics[width=0.445\textwidth]{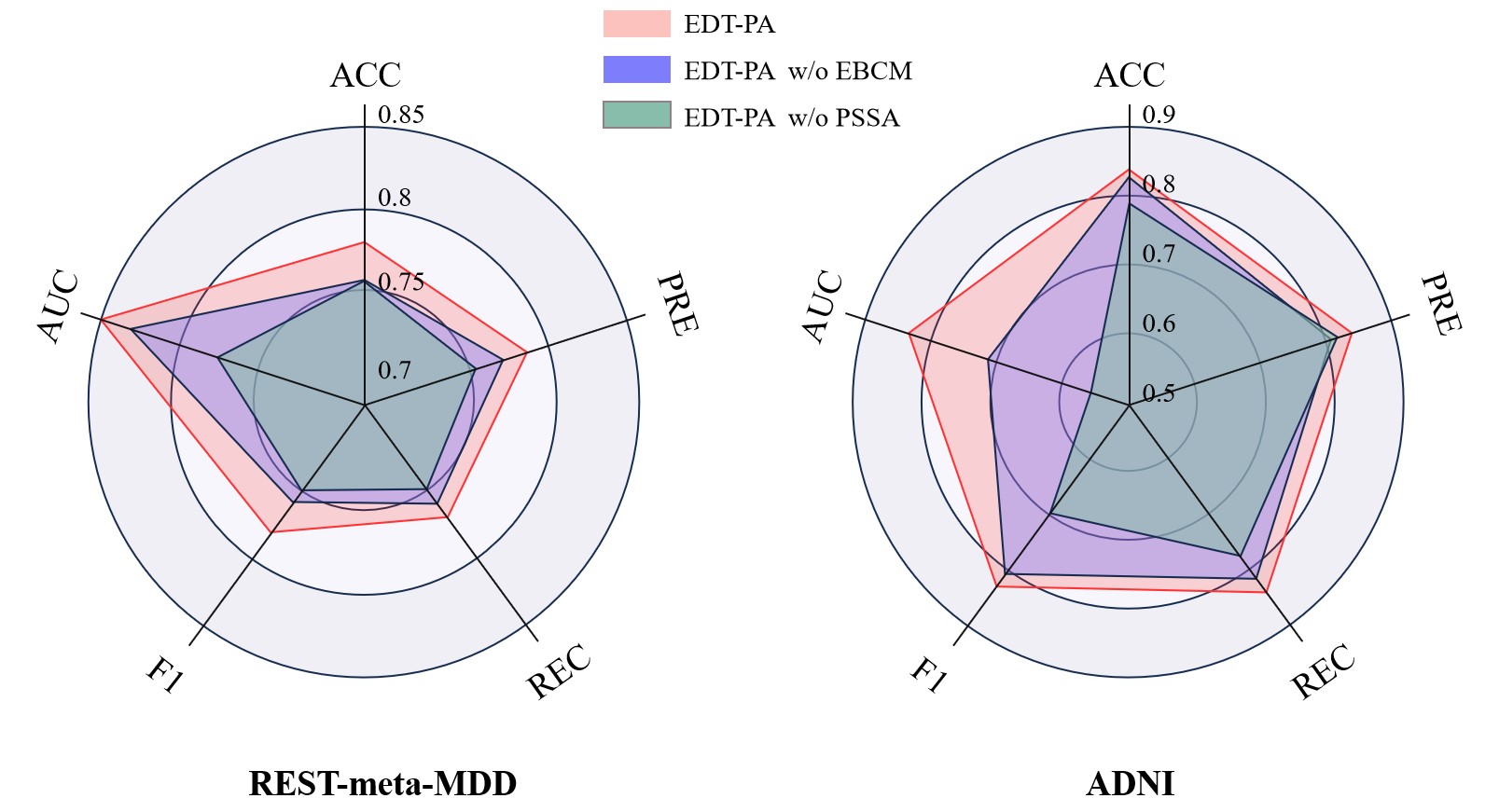}
    \caption{Ablation study of EDT-PA on two public datasets.}
    \label{fig: abstudy}
    \vspace{-0.325cm}
\end{figure}

Two ablation studies are conducted to evaluate the contributions of the EBCM and PSSA modules in EDT-PA. As illustrated in Fig.~\ref{fig: abstudy}, five evaluation metrics (ACC, PRE, REC, F1, and AUC) are compared on the REST-meta-MDD and ADNI datasets by selectively removing each component. In the w/o EBCM setting, the topological evolution process is disabled, and the original brain graph is directly used without diffusion-based enhancement. In the w/o PSSA case, the structure-function alignment mechanism is omitted.

The complete EDT-PA framework consistently delivers the best overall performance across both datasets. Excluding EBCM results in significant reductions in ACC and F1 score, especially on the ADNI dataset, underscoring the critical role of high-order structural modeling. Excluding PSSA mainly degrades AUC and REC, indicating that structure-function misalignment weakens the model’s ability to integrate modality-specific patterns. These results underscore the complementary roles of EBCM and PSSA: the former enhances structural abstraction and evolution, while the latter facilitates modality-aware fusion. Their joint integration is critical for robust and generalizable performance in multimodal brain connectome analysis.



\begin{figure*}[!t]
\centering
    \includegraphics[width=0.95\textwidth]{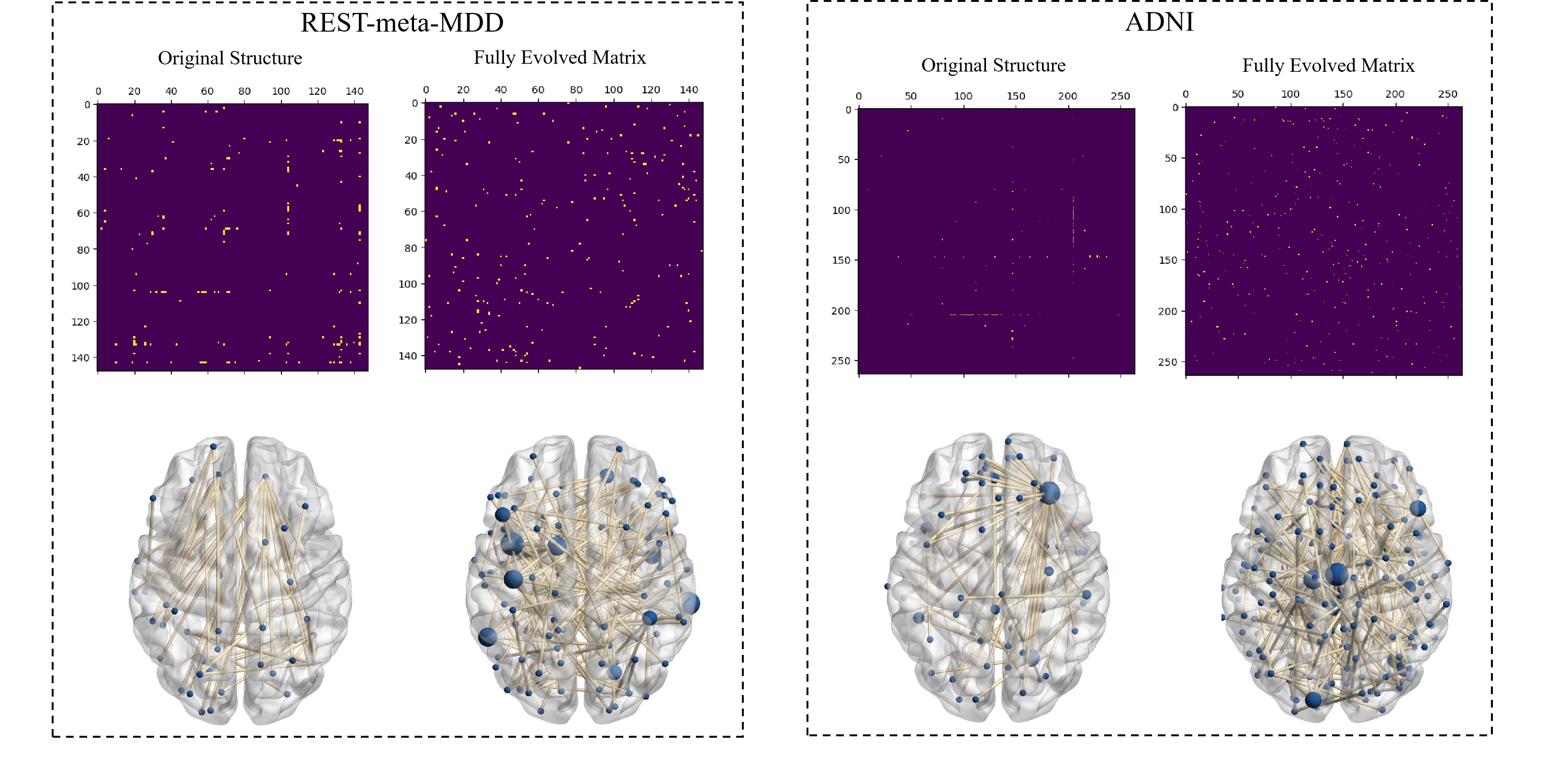}
    \caption{The group difference of original structural matrices and fully evolved matrices in two classification tasks. The connections with
significant difference (p-value $<$ 0.005) are denoted by yellow points in the matrices. The size of a node in the brain network is related to its degree, where a higher degree results in a larger node size.
}
    \label{fig: significance}
\end{figure*}
\vspace{-0.15cm}
\subsection{4.5\quad Explanation Study}


\begin{table}[htb]
	\caption{The Top-10 significant regions detected by EDT-PA in REST-meta-MDD and ADNI dataset.}
				\centering
		\resizebox{1.0\linewidth}{0.12\textheight}{
        \renewcommand{\arraystretch}{1.2}
	\begin{tabular}{cccc}
   \hline
             \multicolumn{4}{c}{\textbf{REST-meta-MDD}} \\ \hline

	\hline
	\multicolumn{1}{c}{Index} & \multicolumn{1}{c
	}{Label} & \multicolumn{1}{c}{Index} & \multicolumn{1}{c}{Label} \\ 
	\hline
		1     &\color[HTML]{9A0000}{Supp\textunderscore Motor\textunderscore Area.L1}          &6     &\color[HTML]{9A0000}{Cingulum\textunderscore Post.L4}\\

		2     & Temporal\textunderscore Inf.L3   
        &7      &\color[HTML]{9A0000}{Occipital\textunderscore Mid.R2   }                         \\

		3     & \color[HTML]{9A0000}{Cingulum\textunderscore Post.L1}      
        &8     & \color[HTML]{9A0000}{Calcarine.L1}   
        \\
		4     &  Occipital\textunderscore Sup.L1
                & 9                         &Frontal\textunderscore Inf\textunderscore Tri.R1  
        \\
		5     &Cerebellum\textunderscore Crus1.R2      
        
        & 10                        &\color[HTML]{9A0000}{ Rolandic\textunderscore Oper.R1}   \\
         \hline  

          \hline 
          \multicolumn{4}{c}{\textbf{ADNI}} \\ \hline
	\multicolumn{1}{c}{Index} & \multicolumn{1}{c
	}{Label} & \multicolumn{1}{c}{Index} & \multicolumn{1}{c}{Label} \\ 
	\hline
		1     &\color[HTML]{9A0000}{S\textunderscore circular\textunderscore insula\textunderscore sup}          &6     &\color[HTML]{9A0000}{G\textunderscore Ins\textunderscore lg\textunderscore and\textunderscore S\textunderscore cent\textunderscore ins}\\

		2     & \color[HTML]{9A0000}{G\textunderscore temporal\textunderscore middle}   
        &7      &\color[HTML]{9A0000}{S\textunderscore front\textunderscore inf    }                         \\

		3     & \color[HTML]{9A0000}{G\textunderscore and\textunderscore S\textunderscore cingul-Mid-Ant}      
        &8     & \color[HTML]{9A0000}{S\textunderscore parieto\textunderscore occipital}   
        \\
		4     &  S\textunderscore interm\textunderscore prim-Jensen 
                & 9                         &\color[HTML]{9A0000}{S\textunderscore postcentral    }
        \\
		5     &\color[HTML]{9A0000}{G\textunderscore precentral}       
        
        & 10                        & S\textunderscore suborbital     \\
         \hline  
	\end{tabular}
}
\label{tab:vis}
\vspace{-0.45cm}
\end{table}
In brain disorder diagnostics, EDT-PA achieves optimal integration of SC and FC while deriving the fully evolved connectivity matrix $\hat{A}$. To evaluate the discriminative power of $\hat{A}$, a statistical significance analysis was performed across two independent datasets, demonstrating its effectiveness in brain disease diagnostic tasks.
The original structural brain network $A$ and $\hat{A}$  are divided according to the health status of the subjects, followed by a significance difference analysis across the data from the different subgroups. The experimental results are shown in Fig.~\ref{fig: significance}. Compared to the original brain structural matrix, $\hat{A}$  exhibits more discriminative connections, demonstrating its ability to capture higher-order dependencies in the brain. This indicates that $\hat{A}$ can precisely identify critical connections related to brain disorders.

A more intriguing discovery is that $\hat{A}$ is predominantly concentrated in several distinct brain regions. To further explore how $\hat{A}$ captures biomarkers associated with brain diseases, the names of the top-10 brain regions with the most significant differential connections are visualized in Table~\ref{tab:vis}. The brain regions associated with the disease are highlighted in red.
 EDT-PA identified several key brain regions associated with depression, including areas in the motor, cingulate, occipital, and frontal regions, as well as the Rolandic operculum. These regions show alterations in connectivity pattern, affecting visual processing, emotional regulation, and cognitive functions~\cite{zhang2024effect, taylor2013fiber,sun2022comparative,lai2017functional,liu2020cacna1c,trombello2022neural}. Using the ADNI dataset, EDT-PA identified several important brain areas, including regions in the insula, temporal lobe, cingulate cortex, frontal lobe, and occipital lobe, as well as the precentral gyrus. These regions are particularly linked to impairments in cognitive functions, emotional regulation, and motor abilities, which are essential for understanding the progression of Alzheimer’s disease~\cite{wan2014identifying,He2009Voxel-based,Pievani2017Coordinate-Based,Foundas1997Atrophy}.

\section{5 \quad Conclusion}
In this study, we propose an end-to-end graph learning framework for analyzing brain connectivity and classifying brain disorders.
The EDT-PA efficiently combines structural and functional brain networks by employing a novel optimal transport method. The framework dynamically captures high-order dependencies and models complex interactions within brain regions, providing a robust approach for disease-specific connectome pattern identification. Extensive evaluations on two real-world datasets demonstrate that EDT-PA outperforms state-of-the-art methods in both classification accuracy and robustness, underscoring its potential to identify disease-specific biomarkers and improve brain disorder diagnostics. This work offers a promising approach for modeling complex brain networks, significantly advancing neuroimaging-based disease diagnosis.

\fi
\setlength{\leftmargini}{20pt}
\makeatletter\def\@listi{\leftmargin\leftmargini \topsep .5em \parsep .5em \itemsep .5em}
\def\@listii{\leftmargin\leftmarginii \labelwidth\leftmarginii \advance\labelwidth-\labelsep \topsep .4em \parsep .4em \itemsep .4em}
\def\@listiii{\leftmargin\leftmarginiii \labelwidth\leftmarginiii \advance\labelwidth-\labelsep \topsep .4em \parsep .4em \itemsep .4em}\makeatother

\setcounter{secnumdepth}{0}
\renewcommand\thesubsection{\arabic{subsection}}
\renewcommand\labelenumi{\thesubsection.\arabic{enumi}}

\newcounter{checksubsection}
\newcounter{checkitem}[checksubsection]

\newcommand{\checksubsection}[1]{%
  \refstepcounter{checksubsection}%
  \paragraph{\arabic{checksubsection}. #1}%
  \setcounter{checkitem}{0}%
}

\newcommand{\checkitem}{%
  \refstepcounter{checkitem}%
  \item[\arabic{checksubsection}.\arabic{checkitem}.]%
}
\newcommand{\question}[2]{\normalcolor\checkitem #1 #2 \color{blue}}
\newcommand{\ifyespoints}[1]{\makebox[0pt][l]{\hspace{-15pt}\normalcolor #1}}
\bibliography{ref}

\end{document}